# A Note on Efficient Computation of All Abelian Periods in a String

M. Crochemore[a,b], C. S. Iliopoulos[a,c], T. Kociumaka[d], M. Kubica[d],
J. Pachocki[d], J. Radoszewski[d,*], W. Rytter[d,e,1], W. Tyczyński[d], T. Waleń[f,d]

[a] *King's College London, London WC2R 2LS, UK*
[b] *Université Paris-Est, France*
[c] *Digital Ecosystems & Business Intelligence Institute, Curtin University of Technology, Perth WA 6845, Australia*
[d] *Faculty of Mathematics, Informatics and Mechanics, University of Warsaw, ul. Banacha 2, 02-097 Warsaw, Poland*
[e] *Dept. of Math. and Informatics, Copernicus University, ul. Chopina 12/18, 87-100 Toruń, Poland*
[f] *Laboratory of Bioinformatics and Protein Engineering, International Institute of Molecular and Cell Biology in Warsaw, Poland*

**Abstract**

We derive a simple efficient algorithm for Abelian periods knowing all Abelian squares in a string. An efficient algorithm for the latter problem was given by Cummings and Smyth in 1997. By the way we show an alternative algorithm for Abelian squares. We also obtain a linear time algorithm finding all "long" Abelian periods. The aim of the paper is a (new) reduction of the problem of all Abelian periods to that of (already solved) all Abelian squares which provides new insight into both connected problems.

*Keywords:* algorithms, Abelian period, Abelian square

## 1. Introduction

We present an efficient reduction of the Abelian period problem to the Abelian square problem. For a string of length $n$ the latter problem was solved in $O(n^2)$ by Cummings and Smyth [7]. The best previously known algorithms for the Abelian periods, see [12], worked in $O(n^2 m)$ time (where $m$ is the alphabet size) which for large $m$ is $O(n^3)$. Our algorithm works in $O(n^2)$ time, independently of the alphabet size. As a by-product we obtain an alternative

*Corresponding author. Tel.: +48-22-55-44-484, fax: +48-22-55-44-400.
*Email addresses:* maxime.crochemore@kcl.ac.uk (M. Crochemore),
c.iliopoulos@kcl.ac.uk (C. S. Iliopoulos), kociumaka@mimuw.edu.pl (T. Kociumaka),
kubica@mimuw.edu.pl (M. Kubica), pachocki@mimuw.edu.pl (J. Pachocki),
jrad@mimuw.edu.pl (J. Radoszewski), rytter@mimuw.edu.pl (W. Rytter),
w.tyczynski@mimuw.edu.pl (W. Tyczyński), walen@mimuw.edu.pl (T. Waleń)
[1]The author is supported by grant no. N206 566740 of the National Science Centre.

$O(n^2)$ time algorithm finding all Abelian squares and an $O(n)$ time algorithm finding a compact representation of all Abelian periods of length greater than $n/2$, in particular, the shortest such period.

Abelian squares were first studied by Erdös [11], who posed a question on the smallest alphabet size for which there exists an infinite Abelian-square-free string. An example of such a string over five-letter alphabet was given by Pleasants [16] and afterwards the best possible example over four-letter alphabet was shown by Keränen [13].

Quite recently there have been several results on Abelian complexity in words [1, 4, 8, 9, 10] and partial words [2, 3] and on Abelian pattern matching [5, 14, 15]. Abelian periods were first defined and studied by Constantinescu and Ilie [6].

We say that two strings are (commutatively) equivalent, and write $x \equiv y$, if one can be obtained from the other by permuting its symbols. In other words, the Parikh vectors $\mathcal{P}(x), \mathcal{P}(y)$ are equal, where the Parikh vector gives frequency of each symbol of the alphabet in a given string. Parikh vectors were introduced already in [6] for this problem.

A string $w$ is an *Abelian k-power* if $w = x_1 x_2 \ldots x_k$, where

$$x_1 \equiv x_2 \equiv \ldots \equiv x_k$$

The size of $x_1$ is called the *base* of the $k$-power. In particular $w$ is an Abelian square if and only if it is an Abelian 2-power.

A string $x$ is an Abelian factor of $y$ if $\mathcal{P}(x) \leq \mathcal{P}(y)$, that is, each element of $\mathcal{P}(x)$ is smaller than the corresponding element of $\mathcal{P}(y)$. The pair $(i, p)$ is an *Abelian period* of $w = w[1, n]$ if and only if $w[i+1, j]$ is an Abelian $k$-power with base $p$ (for some $k$) and $w[1, i]$ and $w[j+1, n]$ are Abelian factors of $w[i+1, i+p]$, see Fig. 1. Here $p$ is called the *length* of the period.

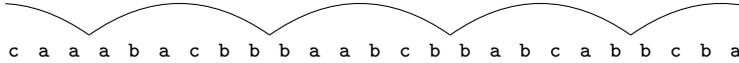

c a a a b a c b b b a a b c b b a b c a b b c b a

Figure 1: A word of length 25 with an Abelian period $(i = 3, p = 6)$. This period implies two Abelian squares: `abacbbbaabcb` and `baabcbbabcab`.

In Section 2 we introduce two auxiliary tables that we use in computing Abelian squares and powers. Next in Section 3 we show new $O(n^2)$ time algorithms for all Abelian squares and all Abelian periods in a string and a reduction between these problems.

Finally in Section 4 we present an $O(n)$ time algorithm finding a compact representation of all "long" Abelian periods. Define

$$MinLong(i) = \min\{p > n/2 \,:\, (i, p) \text{ is an Abelian period of } w\}.$$

If no such $p$ exists, we set $MinLong(i) = \infty$. All long Abelian periods are of the form $(i, p)$ where $p \geq MinLong(i)$, the table $MinLong$ is a *compact* $O(n)$ space representation of potentially quadratic set of long Abelian periods.

## 2. Auxiliary tables

Let $w$ be a string of length $n$. Assume its positions are numbered from 1 to $n$, $w = w_1 w_2 \ldots w_n$. By $w[i, j]$ we denote the factor of $w$ of the form $w_i w_{i+1} \ldots w_j$. Factors of the form $w[1, i]$ are called prefixes of $w$ and factors of the form $w[i, n]$ are called suffixes of $w$.

We introduce the following table:

$$head(i, j) = \text{minimum } k \text{ such that } \mathcal{P}(w[i, j]) \leq \mathcal{P}(w[j + 1, j + k]).$$

If no such $k$ exists, we set $head(i, j) = \infty$, and if $j < i$, we set $head(i, j) = 0$. In the algorithm below we actually compute a slightly modified table $head'(i, j) = j + head(i, j)$.

**Example 1.** *For the infinite Fibonacci word $\mathcal{F} = abaababaabaababaababaa\ldots$ the first several values of the table $head(1, i)$ are:*

| $i$ | 1 | 2 | **3** | 4 | **5** | **6** | 7 | **8** | 9 | **10** | **11** | ... |
|---|---|---|---|---|---|---|---|---|---|---|---|---|
| $\mathcal{F}[i]$ | $a$ | $b$ | $a$ | $a$ | $b$ | $a$ | $b$ | $a$ | $a$ | $b$ | $a$ | ... |
| $head(1, i)$ | *2* | *3* | **3** | *5* | **5** | **6** | *8* | **8** | *10* | **10** | **11** | ... |

*We have here Abelian square prefixes of lengths 6, 10, 12, 16, 20, 22.*

We show how to compute the $head'$ table in $O(n^2)$ time. The computation is performed in row-order of the table using the fact that it is non-decreasing:

**Observation 2.** *For any $1 \leq i \leq j < n$, $head'(i, j) \leq head'(i, j + 1)$.*

We assume that the alphabet of $w$ is $\Sigma = \{1, 2, \ldots, m\}$ where $m \leq n$. For a Parikh vector $Q$, by $Q[i]$ for $i = 1, 2, \ldots, m$ we denote the number of occurrences of the letter $i$. For two Parikh vectors $Q$ and $R$, we define their *Parikh difference*, denoted as $Q - R$, as a Parikh vector: $(Q - R)[i] = Q[i] - R[i]$.

In the algorithm we store the difference $\Delta_j = \mathcal{P}(y_j) - \mathcal{P}(x_j)$ of Parikh vectors of

$$x_j = w[i, j] \quad \text{and} \quad y_j = w[j + 1, k]$$

where $k = head'(i, j)$. Note that $\Delta_j[a] \geq 0$ for any $a = 1, 2, \ldots, m$.

Assume we have computed $head'(i, j - 1)$ and $\Delta_{j-1}$. When we proceed to $j$, we move the letter $w[j]$ from $y$ to $x$ and update $\Delta$ accordingly. Thus at most one element of $\Delta$ might have dropped below 0. If there is no such element, we conclude that $head'(i, j) = head'(i, j - 1)$ and that we have obtained $\Delta_j = \Delta$. Otherwise we keep extending $y$ to the right with new letters and updating $\Delta$ until all its elements become non-negative. We obtain the following algorithm Compute-*head*.

**Lemma 3.** *The head table can be computed in $O(n^2)$ time.*

PROOF. The time complexity of the algorithm Compute-*head* is $O(n^2)$. Indeed, the total number of steps of the while-loop for a fixed value of $i$ is $O(n)$, since each step increases the variable $k$. □

We also use the following *tail* table that is analogical to the *head* table:

$$tail(i, j) = \text{minimum } k \text{ such that } \mathcal{P}(w[i, j]) \leq \mathcal{P}(w[i - k, i - 1]).$$

```
Algorithm Compute-head(w)
    for i := 1 to n do
        Δ := (0, 0, ..., 0);
        Δ[w[i]] := 1; {Boundary condition}
        k := i;
        for j := i to n do
            Δ[w[j]] := Δ[w[j]] − 2;
            while (k < n) and (Δ[w[j]] < 0) do
                k := k + 1;
                Δ[w[k]] := Δ[w[k]] + 1;
            if Δ[w[j]] < 0 then k := ∞;
            head'(i, j) := k;  head(i, j) := head'(i, j) − j;
```

## 3. Abelian squares and Abelian periods

In this section we show how Abelian periods can be inferred from Abelian squares in a string.

Define by $maxpower(i, p)$ the maximal size of a prefix of $w[i, n]$ which is an Abelian $k$-power with base $p$ (for some $k$). Define $square(i, p) = 1$ if and only if $maxpower(i, p) \geq 2p$. Cummings and Smyth [7] compute an alternative table $square'(i, p)$, such that $square'(i, p) = 1$ if and only if $w[i − p + 1, i + p]$ is an Abelian square. These tables are clearly equivalent:

$$square(i, p) = 1 \Leftrightarrow square'(i + p − 1, p) = 1.$$

The $maxpower(i, p)$ table can be computed from the $square(i, p)$ table in linear time using a simple dynamic programming recurrence:

$$maxpower(i, p) = \begin{cases} 0 & \text{if } n - i < p - 1 \\ p + square(i, p) \cdot maxpower(i + p, p) & \text{otherwise.} \end{cases} \quad (1)$$

An alternative $O(n^2)$ time algorithm for computing the table $square(i, p)$ for a string $w$ of length $n$ is a consequence of the following observation, see also Example 1.

**Observation 4.** $square(i, p) = 1 \Leftrightarrow head(i, i + p − 1) = p$.

**Theorem 5.** *All Abelian squares in a string of length $n$ can be computed in $O(n^2)$ time.*

The following observation provides a constant-time condition for checking an Abelian period.

**Observation 6.** $(i, p)$ *is an Abelian period of* $w$ *if and only if*

$$p \geq head(1, i), tail(j, n)$$

where $j = i + 1 + maxpower(i + 1, p)$.

We conclude with the following algorithm for computing Abelian periods. In the algorithm we use our alternative version of computing the table *square* from *head*, since the latter table is computed anyway (instead of that Cummings and Smyth's algorithm can be used for Abelian squares).

---

**Algorithm** Compute-Abelian-Periods
    Compute $head(i, j)$, $tail(i, j)$ using algorithm Compute-*head*;
    Initialize the table *maxpower* to zero table;
    **for** $p := 1$ **to** $n$ **do**
      **for** $i := n$ **downto** $1$ **do**
        **if** $i \leq n - p + 1$ **then**
          $maxpower(i, p) := p$;
          **if** $head(i, i + p - 1) = p$ **then**
            $maxpower(i, p) := p + maxpower(i + p, p)$;
    **for** $i := 0$ **to** $n - 1$ **do**
      **for** $p := 1$ **to** $n - i$ **do**
        $j := i + 1 + maxpower(i + 1, p)$;
        **if** $(p \geq head(1, i))$ **and** $(p \geq tail(j, n))$ **then**
          Report an Abelian period $(i, p)$;

---

**Theorem 7.** *All Abelian periods of a string of length* $n$ *can be computed in* $O(n^2)$ *time.*

### 4. Long Abelian periods

In this section we show how to compute the table $MinLong(i)$, see the example in the table below.

| $i$ | 0 | 1 | 2 | 3 | 4 | 5 | 6 | 7 | 8 | 9 | 10 | 11 | 12 | 13 |
|---|---|---|---|---|---|---|---|---|---|---|---|---|---|---|
| $w[i]$ | | $c$ | $a$ | $a$ | $b$ | $b$ | $c$ | $a$ | $b$ | $b$ | $c$ | $a$ | $a$ | $a$ |
| $MinLong(i)$ | 7 | 7 | 9 | 8 | 7 | 7 | $\infty$ | $\infty$ | $\infty$ | $\infty$ | $\infty$ | $\infty$ | $\infty$ | $\infty$ |

For a non-decreasing function $f : \{1, 2, \ldots, n+1\} \to \{-\infty\} \cup \{1, 2, \ldots, n+1\}$ define the function

$$\hat{f}(i) = \min\{j : f(j) > i\}.$$

If the minimum is undefined then we set $\hat{f}(i) = \infty$.

**Observation 8.** *Let $f$ be a function non-decreasing and computable in constant time. Then all the values of $\hat{f}$ can be computed in linear time.*

**Theorem 9.** *A compact representation of all long Abelian periods can be computed in linear time.*

PROOF. Let us take $f(j) = j - tail(j,n)$. This function is non-decreasing, see also Observation 2. Then for $i < \frac{n}{2}$ we have:

$$MinLong(i) \;=\; \max\left\{\left\lfloor\frac{n}{2}\right\rfloor + 1,\; head(1,i),\; \hat{f}(i) - i - 1\right\}$$

and otherwise $MinLong(i) = \infty$, see also Fig. 2.

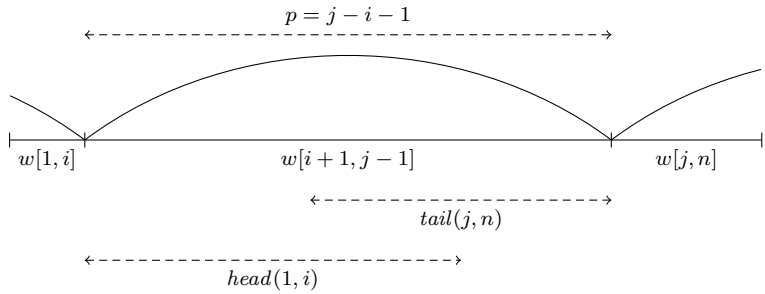

Figure 2: A schematic view of a long Abelian period: $p > \frac{n}{2}$, $p \geq head(1,i)$, $tail(j,n)$.

Hence the computation of *MinLong* table is reduced to linear time algorithm for $\hat{f}$ and the conclusion of the theorem follows from Observation 8. □